\documentclass[12pt]{iopart}
\usepackage{amsfonts}
\usepackage{amssymb}
\usepackage{mathrsfs}
\usepackage{bm}
\usepackage{graphicx}%
\usepackage{mathbbol}
\newcommand{\beq}{\begin{equation}}
\newcommand{\eeq}{\end{equation}}

\newcommand{\beqa}{\begin{eqnarray}}
\newcommand{\eeqa}{\end{eqnarray}}
\def\ra{\rangle}
\def\la{\langle}

\begin{document}
\title{Symmetries and time operators}
\author{G. C. Hegerfeldt}
\address{Institut f\"ur Theoretische Physik, Universit\"at G\"ottingen,
Friedrich-Hund-Platz 1, 37077 G\"ottingen, Germany, and}
\address{Max Planck Institute for the Physics of Complex Systems,
  N\"othnitzer Str. 38, 01187 Dresden, Germany}  
\author{J. G. Muga}
\address{Departamento de Qu\'{\i}mica F\'{\i}sica, Universidad del
Pa\'{\i}s Vasco, Apartado 644, 48080 Bilbao, Spain, and}
\address{Max Planck Institute for the Physics of Complex Systems,
  N\"othnitzer Str. 38, 01187 Dresden, Germany}  
%
\begin{abstract}
All covariant time operators with normalized
probability distribution are derived. Symmetry criteria are invoked to
arrive  at a unique expression for a given Hamiltonian. As an
application, a well known result
for the arrival time distribution of a free particle is
generalized and extended. Interestingly, the resulting arrival time
distribution operator  is connected to a particular, 
positive, quantization of the classical current. For particles in a
potential we also introduce and study the 
notion of conditional arrival-time distribution.
\end{abstract}
%
%
%
\section{Introduction}
In spite of a famous footnote by Pauli \cite{Pauli} on the purported
non-existence of a time operator in quantum mechanics, physically
motivated questions relating to time occur  quite naturally in 
the form of arrival times or of time durations. These become
interesting when the extension of the wavefunction comes into play, and
a quantum mechanical description by operators is not at all obvious.
Also clock time operators have been investigated in the literature.
For reviews of several aspects of these difficulties and of some more recent
results see, e.g. \cite{b1,b2}. As quite recently stressed 
by us and  Mu\~{n}oz in \cite{HeMu1}, for a given system 
there is no single time operator. Indeed, there may be 
many,  corresponding to different observables and
measuring devices.


There are two main groups of physically relevant time
operators. These  are associated with time durations and time
instants, respectively. An example of a duration is the dwell time of a particle in
a region of space.  The corresponding operator commutes with the
Hamiltonian since the duration  does not depend on a particular 
reference instant \cite{dwell}. In the other group, the time
observables are covariant in the sense that when the   preparation time
is shifted they are shifted by the same amount, either forward
(quantum clocks) \cite{clocks} or backward (event times recorded with a
stopwatch, e.g. time of arrival), and they are conjugate to the
Hamiltonian. 

The measurement of a clock time operator $\hat T$ is supposed to yield, up
to a constant, the parametric time. More precisely, $\hat T$ should
satisfy  $\la\psi_t|\hat{T}|\psi_t\ra = t + \la\psi_0|\hat{T}|\psi_0\ra$. 
Time-of arrival operators are mathematically similar to clock time
operators, but the physics 
is different. Some of their properties are motivated by the behaviour of a classical particle, and in any case may be put at test experimentally. A free particle in one dimension must, during
its motion, arrive at a given point with certainty, and  it must also
arrive in three dimensions at an infinitely extended plane. On a
half-line, a free particle is reflected at the end point and may therefore
arrive twice at another point. In this case it is reasonable to
investigate {\em first} (or second) arrivals at a point \cite{HeMu1}. The most difficult case is 
a particle interacting with an external
potential \cite{pot1,pot2,pot3,Galapon06,Galapon08,Galapon}. The particle can be
deflected 
and, in one dimension, reflected so that it may, with a non-zero
probability, never arrive at a particular position. In such a
case it is then 
natural to ask for the probability distribution of first arrival at time $t$,
given that the particle is arriving at all, i.e. to ask for the
{\it conditional} first-arrival probability distribution, which is, by
definition,  normalized to 1. 
In \cite{HeMu1}  the most general form of covariant 
and normalized time observables was announced and a proof of a special
case was indicated. The general proof of this
result will be given in this paper. Reference \cite{HeMu1} also gave
applications to arrival times for a free particle on a 
half line and to Lyapunov operators in quantum mechanics. Based on physically
motivated postulates pertaining to 
arrivals at a plane in three dimensions,  \cite{Kij} and \cite{Werner}
derived a time-of-arrival operator  for a free particle in three
dimensions coming in from one side of the plane. Among  the postulates
were invariance of the time operator under Galilei transformations,
which leave invariant the plane in question, and minimal means-square
deviation. This result will be generalized in the present paper and
the underlying conditions will be weakened. In particular,  Galilei
invariance will not be needed. 

The plan of the paper is as follows.  In Section \ref{general} we study
covariant and normalized time observables and explicitly  spell out
their most general form, with the detailed proof referred to the
Appendix. In Section \ref{uniqueness} we analyze physical conditions,
in particular symmetry conditions, under which time observables
become unique, and discuss applications to  clock-time operators. In  Section
\ref{arrival} we give applications to time-of-arrival operators for free
particles in one and three dimensions. In particular, the result of
\cite{Kij,Werner}  will be generalized.  The
restriction on the incoming direction will be lifted and 
the above Galilei invariance will be replaced
by invariance under translations which leave the plane in question
invariant. It will be shown that the resulting arrival time
distribution and operator are related to a particular, 
positive, quantization of the classical current. Due to the backflow
effect \cite{backflow} the usual
quantization in the form of the quantum mechanical probability does
not have the necessary positivity properties.
In  Section \ref{cond} we study the notion of first arrivals in the
presence of a potential and the
associated probability distribution operator, which in general is not
normalized to 1. Normalizing by the total arrival probability for a
given state one obtains the conditional
arrival probability distribution which is normalized, but, by construction, is
not bilinear in the wavefunction.  Alternatively, to keep bilinearity we  consider 
normalization by means of an operator and the resulting distribution
operators. A one-dimensional example for a 
particle in a  potential symmetric about the origin and arrivals at
the origin is discussed. In the Appendix
mathematical fine points such as regularity properties are discussed.
\section{Clock time and time-of-arrival operators}\label{general}
Clock time operators can be associated with
a quantum clock whose average measures the progressing parametric
time. A good clock would also minimize the variance in order to
estimate the time as accurately as possible with a finite number of measurements.

For a clock observable, let the probability of finding the
measured time in the interval $(\infty, \tau)$ for a state $|\psi\rangle$ be given by the expectation of an operator $\hat{F}_\tau$,
which plays the role of a cumulative probability operator, so that $\hat{F}_\infty = \Eins$. 
The derivative $d\langle \psi | \hat{F}_\tau|\psi \rangle/d\tau$ 
is the corresponding temporal probability distribution and it is normalized to
1. For a momentum measurement the  analogous operator would be a
projector, while here one just has  $0\leq 
\hat{F}_\tau \leq 1 ~$ with selfadjoint $\hat{F}_\tau$. The collection
of $\hat F_\tau$'s yields a positive operator-valued measure.  We
define the probability distribution operator $\hat \Pi_\tau$ and the
time operator $\hat T$ by
\beqa
\hat{\Pi}_\tau &\equiv& \frac{d}{d\tau} \hat{F}_\tau~,
\nonumber\\
\label{eq1.2}
\hat{T} &\equiv& \int d\tau\, \tau\, \hat{\Pi}_\tau~.
\eeqa
For details on the derivative see the Appendix.  The mean value
of observed time can be written as 
\beq
\label{eq1.3}
\langle \psi| \int d\tau\, \tau\, \hat{\Pi}_\tau | \psi \rangle 
\equiv \langle \psi | \hat{T} | \psi \rangle~.
\eeq
The second moment, if it exists, is given by $\langle \psi| \int
d\tau\,\tau^2\, \hat{\Pi}_\tau\,  
|\psi \rangle$
and similarly for higher moments.

Covariance of a clock time operator  with respect to ordinary
(parametric) time means that the probabilities of finding the
measured time in the  interval $(- \infty, \tau)$ for state $ | \psi_0
\rangle$ and in $(- \infty, \tau + t)$ for $|\psi_{t} \rangle$ coincide. This
implies
\beq
\label{eq1.7a}
\hat{F}_\tau = e^{-{\rm i} \hat{H} \tau/\hbar} \hat{F}_0\, e^{{\rm i} \hat{H}\tau/\hbar}.
\eeq
{}From this one obtains by differentiation (see the Appendix for details)
\begin{eqnarray}
\label{eq1.7bc}
\hat{\Pi}_0 &=&  \frac{-{\rm i}}{\hbar}[\hat H, \hat F_0],
\\
\label{eq1.7b}
\hat{\Pi}_\tau &=&
e^{-{\rm i} \hat{H} \tau/\hbar}\,\hat{\Pi}_0\, e^{{\rm i} \hat{H}\tau/\hbar}~.
\end{eqnarray}
Note that
$\hat \Pi_0$ and $\hat \Pi_\tau$ are in general not operators on the
Hilbert space but  only bilinear forms evaluated
between normalizable vectors from the domain of $\hat H$.  
Since $\la\psi|\hat{F}_\tau|\psi\ra$ is non-decreasing,
$\hat{\Pi}_\tau$ is a positive bilinear form. 
An expression like $\la E| \hat{\Pi}_0|E'\ra$ has to be understood as a
distribution. Since the diagonal $E=E'$ has measure 0 it is no contradiction
that  (\ref{eq1.7bc}) formally gives 0 on 
the diagonal while explicit examples may give a non-zero value \cite{HeMu1}.
By a change of variable in (\ref{eq1.2}) one obtains
\begin{equation}
\label{eq1.7c}
e^{{\rm i} \hat{H}t/\hbar}\, \hat{T} \,e^{-{\rm i} \hat{H}t/\hbar} = \hat{T} + t.
\end{equation}
Furthermore, $\hat H$ and $\hat T$ satisfy the canonical commutation
relations as bilinear forms on a suitable domain. 

A superscript $A$ denotes quantities for (first) arrival times at a
given location.
In contrast to clock times, if the particle's state is
shifted in time by $t_0$, it 
should arrive a time $t_0$ earlier, and the temporal probability distribution
should be shifted by $t_0$ to earlier times.    
Thus the analog of the cumulative probability operator in
(\ref{eq1.7a}) and the probability density operator
must now satisfy   
\begin{eqnarray}
\label{eq5.7a}
\hat{F}^A_t &=& e^{{\rm i} \hat{H} t/\hbar} \hat{F}^A_0\, e^{-{\rm i} \hat{H}t/\hbar},
\\
\hat{\Pi}^A_0&=&\frac{{\rm i}}{\hbar}[\hat{H},\hat{F}^A_0], \label{eq5.7b} \\
\label{eq5.9a}
\hat{\Pi}^A_{t} &=& e^{{\rm i} \hat Ht/\hbar}\, \hat{\Pi}^A_0 \, e^{-{\rm i}\hat Ht/\hbar}. 
\end{eqnarray}
Again, $\hat \Pi^A_\tau$ is a positive bilinear form. 

In case of a particle in a potential, it may happen that for a given
state the particle arrives at the location with a total probability
less than 1. This means that the integral of $\hat{\Pi}^A_{t}$ is not
unity.  For the average arrival time one can then not use the
left-hand side of the analog of (\ref{eq1.3}) but rather
\beq \label{1.6b}
\langle \psi| \int d\tau\, \tau\, \hat{\Pi}_\tau | \psi \rangle \,/ \,
\langle \psi| \hat{N}| \psi \rangle  \, ,
\eeq
where $\hat{N}\equiv \int d\tau \,  \hat{\Pi}_\tau = \hat F_\infty$.
However, if the total arrival probability is always 1, one can define
an arrival time operator  by
\beqa
\label{eq5.8a}
\hat{T}^A &=& \int d\tau\, \tau\, \hat{\Pi}_\tau \\
&=& \int {dt}\, t \, e^{{\rm i} \hat{H}t/\hbar}\, \hat{\Pi}^A_0 \,
e^{-{\rm i} \hat{H}t/\hbar}.
\eeqa
As seen by a change of integration variable, an arrival time operator 
behaves as the negative of a clock time operator whose probability
distribution operator is given by 
$\hat\Pi_{\tau}=\hat\Pi_{-\tau}^A$, with a cumulative 
probability operator $\hat F_\tau=\Eins -\hat F^A_{-\tau}$.

We now prove the following general result for
covariant clock  operators and arrival time operators with normalized
distributions. This was already announced in
\cite{HeMu1} without proof. In \ref{nondegenerate}
and \ref{degenerate} we present rigorous mathematical details.

Let $\hat H$ be a Hamiltonian with
purely continuous eigenvalues  and
eigenvectors $|E,\alpha\ra$, where the degeneracy index can be assumed
to be a discrete number. Their normalization is taken as 
\begin{equation} \label{eq2.7}
\langle E, \alpha | E^\prime, \alpha^\prime \rangle = \delta_{\alpha
  \alpha^\prime}  \delta(E- E^\prime).
\end{equation}
For simplicity we assume the same degeneracy for each $E$. We consider
any covariant clock time operator which has a probability distribution
normalized to 1 and whose second moment exists for a dense set of vectors.
Then for $t=0$ the probability distribution $\hat{\Pi}_0$ is given  by
\beqa
\hat{\Pi}_0 &=& \frac{1}{2\pi\hbar}\sum_i |b_i\ra\la b_i|
\nonumber\\
&=&\frac{1}{2\pi\hbar}\sum_i\int  dE\,dE^\prime \sum_{\alpha \alpha'}
b_i(E, \alpha)\,|E, \alpha \rangle \langle
E^\prime , \alpha^\prime|\, \overline{b_i (E^\prime, \alpha^\prime)},
\label{eq2.8}
\eeqa
where  $b_i(E, \alpha) \equiv \langle E,\alpha|\,b_i\ra$ 
are any functions satisfying the two conditions
\begin{eqnarray} \label{eq2.9}
&\sum_{i}&  b_i(E, \alpha) \overline{b_i (E, \alpha^\prime)}=
\delta_{\alpha \alpha^\prime},
\\
 \label{eq2.11}
&\sum_{i}& \big|\partial_E \,b_i (E, \alpha)\big|^2 
  \;\; \mbox{integrable over finite 
  intervals.} 
\end{eqnarray}
The same results hold for $\hat \Pi_0^A$  of normalized arrival time
distributions. The complete distributions $\hat \Pi_t$ and $\hat
\Pi_t^A$ and the associated time operators are then given by
(\ref{eq1.2})~-~(\ref{eq1.7b}) and
(\ref{eq5.7a})~-~(\ref{eq5.8a}), respectively.  

{}For a state $|\psi\ra$,  with $\psi (E, \alpha)\equiv \langle E ,
\alpha|\psi \rangle$, one has 
\beqa
\la \psi|\hat T |\psi\ra &=& \int dE \sum_\alpha \overline{\psi(E,\alpha)}
 \, {\rm i}\hbar\,\partial_E \psi(E,\alpha)
\nonumber\\
\label{eq2.10a}
&&\!\!\!\!\!\!\!\!\!\!\!\!\!\!\!\!\!\!\!+ \int dE 
\sum_{\alpha \alpha'} \overline{
  \psi(E,\alpha)}\,\psi(E,\alpha') \sum_ib_i(E,\alpha)\, {\rm i}\hbar\,\partial_E
\overline{b_i(E,\alpha')} 
\eeqa
\begin{equation}\label{eq2.10}
\mbox{second moment} = \hbar^2\!\!\int\!\! dE\,\Big|\partial_E
\sum_\alpha \overline{b_i (E, \alpha)}\, \psi (E, \alpha)
\Big|^2~.~  
\end{equation}

A  mathematically detailed proof will be given in 
\ref{nondegenerate} and  \ref{degenerate}. For simplicity we indicate
here a formal proof for the case that the (continuous) eigenvalues of
$\hat H$ are non-degenerate. We start by constructing the functions $b_i$.
For a given $\hat \Pi_0$, one can choose a maximal set
$\{|g_i\rangle\}$ of vectors satisfying
\beq \label{eq2.6a}
\langle g_i |\hat \Pi_0| g_j \rangle = \delta_{ij} ~.
\eeq
Such a maximal set is easily constructed by the standard Schmidt
orthogonalization procedure. Then a possible set of $|b_i\ra$'s is given by
\beq \label{eq2.6b}
|b_i\ra = \sqrt{2\pi\hbar}~\hat{\Pi}_0| g_i \rangle~.
\eeq
Indeed, from (\ref{eq2.6b}) one has
\beqa \label{eq2.6bb}
\frac{1}{2\pi\hbar} \sum_i \la g_\alpha |b_i\ra \la b_i|g_\beta \ra& =&
  \sum_i \la g_\alpha |\hat \Pi_0|g_i\ra \la g_i|\hat \Pi_0|g_\beta
  \ra \nonumber\\
&=& \sum_i \delta_{\alpha i} \delta_{\beta i} = \delta_{\alpha\beta}. 
\eeqa
Equation (\ref{eq2.6a}) then implies
\beq \label{eq2.6bbb}
\hat \Pi_0 = \frac{1}{2\pi\hbar} \sum_i |b_i\ra \la b_i|,
\eeq
which is (\ref{eq2.8}).  It should
be noted  that the $|b_i\ra$'s in the decomposition of $\hat \Pi_0$ in
(\ref{eq2.8}) are not unique.

The  $|b_i\ra\,$'s need not be normalizable and the $b_i(E)\,$'s could
in principle be distributions. That the   $b_i(E)\,$'s are 
in fact functions will be shown in  \ref{nondegenerate}. They
have to satisfy certain properties in 
order that the total probability is 1 and that the second moment is
finite. Indeed, for given normalized state 
$|\psi \rangle$, the total temporal probability is, with $\psi(E)\equiv \langle
E|\psi \rangle$, 
\beqa
&&\int_{- \infty}^{+ \infty} \frac{dt}{2 \pi} 
 \langle \psi|\, e^{-{\rm i}\hat H t/\hbar}\,
\hat \Pi_0\, e^{{\rm i} \hat H t/\hbar}\,|\psi \rangle 
\nonumber\\
&=& \sum_i \int \frac{dt}{2 \pi\hbar} \int dE\,dE^\prime \, e^{-{\rm i}(E-E^\prime)t/\hbar}
\overline{\psi(E)}\, b_i(E)\,
\overline{b_i(E^\prime)}\, 
\psi(E^\prime)
\nonumber\\
&=& \sum_i  \int dE\, dE^\prime \,
\delta(E-E^\prime)
\overline{\psi (E)}\, b_i (E)\,
\overline{b_i(E^\prime)}\, \psi(E^\prime)
\nonumber \\
&=&
 \sum_i \int dE~\overline{\psi (E)} \sum_i b_i(E)\, \overline{b_i(E)}\,
\psi(E) 
\stackrel{!}{=}1~.\label{eq2.2}
\eeqa
Since $|\psi \rangle$ is arbitrary this implies
\begin{equation} \label{eq2.3}
\sum_i b_i (E) \,\overline{b_i(E)} = 1~.
\end{equation}
The formal use of the $\delta$ function will be justified in 
\ref{nondegenerate}. In a similar way one obtains 
\beqa
\langle \psi |\hat{T}| \psi \rangle &=& \int dE\, \bar{\psi}(E)
\,{\rm i}\hbar\, \psi^\prime (E)
\nonumber\\
\label{eq2.3a}
& +& \int dE\, |\psi (E)|^2 \sum b_i(E)\,{\rm i}\hbar\, \overline{b_i^\prime (E)}~.
\eeqa
Note that $\sum b_i\, \bar{b_i^\prime}$ is purely imaginary, from
(\ref{eq2.3}), and thus vanishes if $b_i$ is real. The second
moment is 
\beqa
&\int & \frac{dt}{2 \pi}\, t^2
\langle \psi |e^{-{\rm i}\hat Ht/\hbar} \hat \Pi_0 e^{{\rm i} \hat Ht/\hbar}|\psi
\rangle
\nonumber\\ 
&=& \hbar \sum_i \int \frac{dt}{2 \pi} \int
dE\,dE^\prime\, \partial_E\, \partial_{E^\prime}\, e^{-{\rm i}(E-E^\prime)t/\hbar}
\overline{\psi(E)}\, b_i(E)
\,\overline{b_i(E^\prime)}\, \psi 
(E^\prime) 
\nonumber\\
& =& \hbar^2 \sum_i \int dE \,\partial_E \left(
\overline{\psi (E)}\, \,b_i(E)\,
\right) \partial_E \left( \overline{b_i(E)}\, \psi (E)  \right)
\nonumber\\
&=& \hbar^2 \int dE \big\{
|\psi^\prime(E)|^2 + \sum_i |b_i^\prime (E)|^2 \, |\psi(E)|^2
\nonumber\\
&~&~~~~~~~~~~~~~+ 2\,{\rm Re}\,
\sum_i \overline{b_i(E)}\, b_i^\prime(E)\, \overline{\psi(E)}\,
\psi^\prime(E) \big\} 
\label{eq2.4}
\eeqa
by (\ref{eq2.3}). This is finite if the contribution from the first
and second term are finite, and for the latter to hold for all
infinitely differentiable functions $\psi(E)$ with compact support one
must have that
$\sum_i |b_i^\prime(E)|^2 $ is integrable over any finite interval.
Again,  differentiability of the $b_i$'s will be shown in the
Appendix, as well as the proof in the case that $\hat H$ has
degenerate eigenvalues. 

Conversely, it is easily checked that any functions $b_i(E,\alpha)$
which satisfy (\ref{eq2.9}) and (\ref{eq2.11}) give rise to a
covariant time operator with normalized probability distribution and
second moment for a dense set of vectors.
\section{Degenerate energy eigenstates: unique  clock time   operators   from  symmetry and minimal variance} \label{uniqueness}  
Clock time operators have a normalized probability distribution, and
this will be assumed in the following.
To reduce the multitude of covariant clock time operators we may require
that the variance $\Delta T$, or mean-square 
deviation $\Delta T^2$, is minimal for all states for which the  second
moment exists. Physically this means that no other time observable  can be
measured more precisely. 
However, requiring minimal variance by itself does not make $\hat{T}$
unique, not even in the case of non-degenerate spectrum of $\hat H$,
as was shown in  Ref. \cite{HeMu1}. But if in addition one
restricts the set of 
functions $b_i$ by symmetry requirements, uniqueness may then follow
from minimality of  $\Delta T$, as will be illustrated now.

We denote the anti-unitary time reversal operator by $\hat\Theta$. In
the x space representation one has $(\hat\Theta\psi)(\bf x) =
\overline{\psi(\bf x)}$. If
the Hamiltonian is time reversal invariant, we may demand 
\begin{equation} \label{eq3.1}
\hat\Theta\, \hat{T}\, \hat\Theta = - \hat{T}
\end{equation}
and similarly for the probability distribution. By  (\ref{eq1.7a}) this
implies 
\begin{equation} \label{eq3.2}
\hat\Theta \,\hat{\Pi}_0\hat \Theta = \hat{\Pi}_0.
\end{equation}
In \cite{HeMu1} it was shown for the non-degenerate eigenvalue
case that time reversal invariance of the Hamiltonian $\hat H$ and
minimal $\Delta T$ together imply that $\hat{T}$ and $\hat\Pi_t$ are
unique  and given by
\beqa
\hat{\Pi}_t &=& \frac{1}{2\pi\hbar}\int dE\,dE'\,
e^{-{\rm i}(E-E')t/\hbar}|E_\Theta \ra\la E_\Theta '|,
\nonumber\\
\hat{T}&=& \int dt\,t\, \hat{\Pi}_t,
\label{eq3.5}
\eeqa
with time reversal invariant $|E_\Theta \rangle$. 

If the Hamiltonian has degenerate continuous eigenvalues 
this is no longer true and one needs additional conditions to obtain
uniqueness, as discussed now.\\[.3cm]
\noindent
{\bf Example 1}: In one dimension we consider
\begin{equation}\label{3.6}
\hat{H} = \hat{P}^2/2m + V(|\hat{X}|)
\end{equation}
and assume a purely continuous spectrum. $\hat{H}$ is invariant under
space reflection, denoted by $\hat{\sigma}$, and under time reversal
$\hat{\Theta}$. We assume $\hat{T}$ and its probability distribution to
have analogous properties. For $\hat{\Pi}_0$ 
this implies 
\begin{equation}\label{3.7}
\hat{\sigma} \hat{\Pi}_0 \hat{\sigma}  =\hat{\Theta} \hat{\Pi}_0
\hat{\Theta} =   \hat{\Theta}
\hat{\sigma}\hat{\Pi}_0\hat{\sigma} \hat{\Theta} = \hat{\Pi}_0~.
\end{equation}
Then, under the assumption of normalizability, the time operator
becomes unique, as shown now. As eigenstates of 
$\hat{H}$ we choose real and either symmetric or 
antisymmetric wavefunctions, denoted by $|E, \pm \rangle$ and
normalized as $\delta (E- E^\prime)$. Then $\hat\Theta |E, \pm \rangle = |E,
\pm \rangle$ and $\hat\sigma |E, \pm \rangle = \pm |E, \pm \rangle$. 
Decomposing the functions $b_{i\alpha}$ in  (\ref{eq2.8}) for
$\hat{\Pi}_0$ into real 
and imaginary parts one finds from $\hat{\Pi}_0 = \frac{1}{2} (\hat{\Pi}_0 +
\hat{\Theta} \hat{\Pi}_0 \hat{\Theta})$ that  (\ref{eq2.8}) can be
written with real functions when one uses $|E, + \rangle$ and $|E, -
\rangle$ as basis. 
Writing again $\hat{\Pi}_0 = \frac{1}{2} (\hat{\Pi}_0 + \hat{\sigma} \hat{\Pi}_0
\hat{\sigma})$ one sees that the mixed $\alpha \alpha^\prime$ terms
cancel so that one can write $\hat{\Pi}_0$ in the form
\begin{equation}\label{eq3.10}
\hat{\Pi}_0 = \frac{1}{2\pi\hbar}\sum_i \sum_{\alpha = \pm} \int
dE\,dE^\prime\, b_{i 
  \alpha}(E )\,|E, \alpha\rangle \langle E^\prime, \alpha |\,b_{i
  \alpha}(E^\prime)
\end{equation}
with real functions $b_{i \alpha}$. Note that  (\ref{eq3.10}) has
no mixed $\alpha \alpha^\prime$ terms and is in this sense
diagonal. The normalization condition (\ref{eq2.9}) is replaced by 
\begin{equation}\label{eq3.11}
\sum_i b_{i \alpha} (E)^2 = 1~~,\qquad \alpha = \pm~.
\end{equation}
Diagonality in $\alpha$ and reality of the functions $b_{i \alpha}$ will be
crucial for determining $b_{i\alpha}$ explicitly. In fact, expanding
the right-hand side of (\ref{eq2.10}) for the second moment, the term
\begin{equation}\label{3.12}
\sum_{i \alpha \alpha^\prime } \int dE\,b_i^\prime (E \alpha)\,
\overline{b_i(E, \alpha^\prime)}\,\, \overline{\psi (E, \alpha)}\,
\psi^\prime (E, \alpha^\prime) + {\rm c.c.}
\end{equation}
 is replaced by
\begin{equation}\label{3.13}
\sum_{i \alpha} \int dE\,b_{i\alpha}^\prime (E)\, b_{i \alpha}(E) \left\{
\overline{\psi (E, \alpha)}\,  \psi^\prime (E, \alpha) + {\rm c.c.}
 \right\}
\end{equation}
and similarly for an analogous term in $\langle\psi| \hat{T}|\psi \rangle$.
Now, one has
\begin{equation}\label{3.14}
\sum_i b^\prime_{i \alpha} (E) b_{i \alpha} (E) =
\frac{1}{2} \partial_E \sum_i b_{i \alpha} (E)^2 = 0
\end{equation}
by  (\ref{eq3.11}). Therefore the only $b_{i \alpha}$ dependent
term in $\Delta T^2$ is 
\begin{equation}\label{eq3.15}
\sum_{i \alpha} \int dE |b^\prime_{i \alpha} (E)|^2 |\psi (E,
\alpha)|^2.
\end{equation}
This becomes minimal if and only if $b_{i \alpha}^\prime \equiv 0$, i.e.
\begin{equation}\label{eq.3.16}
b_{i \alpha}(E) \equiv c_{i \alpha},
\end{equation}
with the real constant $c_{i \alpha}$ satisfying, by
(\ref{eq3.11}),
\begin{equation}\label{eq.3.17}
\sum_i (c_{i \alpha})^2 = 1~, \qquad \alpha = \pm~.
\end{equation}
Inserting this into (\ref{eq3.10}) gives finally
\begin{equation}\label{eq3.18}
\hat{\Pi}_0 = \frac{1}{2\pi\hbar} \int dE~ dE^\prime \left\{
|E, + \rangle \langle E^\prime, +|\, + \, |E, - \rangle, \langle E^\prime, -|
\right\}~.
\end{equation}
From (\ref{eq1.7a}) and (\ref{eq1.2}) one now obtains
the corresponding  operators $\hat \Pi_t$ and $\hat T$. 
Thus, in this one-dimensional example, covariance under time reversal
and reflection plus minimality of $\Delta T$ lead to a unique clock time
operator and a unique associated temporal probability distribution.

Interestingly, when the potential $V$ in (\ref{3.6}) allows the application of
scattering theory, $\hat \Pi_0$ in (\ref{eq3.18}) can 
be re-expressed in terms of the M{\o}ller operators $\hat\Omega_{\pm}$ and 
complex reflection and transmission coefficients, $T(k)$ and $R(k)$. We denote
by $|k\rangle$ momentum eigenstates. Reflection invariance gives
$\hat{\sigma} \hat{\Omega}_{\pm} \hat{\sigma} = \hat{\Omega}_\pm$ and
$R(-k) = R(k),~T(-k) = T(k)$. The interpolation property $\hat{H}
\hat{\Omega}_{\pm} = \hat{\Omega}_{\pm} \hat{H}_0$ shows that the two vectors
\beqa
\label{eq3.20}
&&\big(\hat{\Omega}_+ + \hat{\Omega}_-\big) \big(|k\rangle + | -k
\rangle\big)/\sqrt{2}
\nonumber\\
&&{\rm i} \big(\hat{\Omega}_+ + \hat{\Omega}_-\big) \big(|k\rangle - | -k
\rangle\big)/\sqrt{2}
\eeqa
are eigenvectors of $\hat{H}$ which are symmetric and antisymmetric,
respectively, and invariant under $\hat{\sigma} \hat{\Theta}$ since
$\hat{\Theta} \hat{\Omega}_\pm \hat{\Theta} = \hat{\Omega}_\mp$. Hence
the vectors in (\ref{eq3.20}) are real multiples of $|E, \pm
\rangle$. To determine the respective multiple we use the identities
\begin{equation}\label{eq3.35}
\hat{\Omega}_+ + \hat{\Omega}_- = \hat{\Omega}_- (\hat{S} + 1) =
\hat{\Omega}_+ (1+ \hat{S}^\dagger),
\end{equation}
\begin{equation} \label{eq3.21}
\hat{S}\,|k\rangle = T(k)|k
\rangle + R(k) |-k \rangle~.
\end{equation}
Thus
\begin{equation} \label{eq3.22}
\hat{S}\,\big(|k\rangle \pm | -k\rangle\big) =\big (T(k) \pm
R(k)\big)\big(|k \rangle \pm 
|-k \rangle\big)~.
\end{equation}
This implies that here $T(k) \pm R(k)$ are pure phases. With (\ref{eq3.35})
the two vectors in  (\ref{eq3.20}) become
\beqa
\nonumber
&& \big(1 + \overline{T(k)} + \overline{R(k)}  \big)\,
\hat\Omega_+ \big(|k \rangle + |-k \rangle\big)/\sqrt{2},
\\
\label{eq3.23}
&&{\rm i} \big(1 + \overline{T(k)} + \overline{R(k)}  \big)\,
\hat\Omega_+ \big(|k \rangle - |-k \rangle\big)/\sqrt{2}.
\eeqa
Hence, from normalization one finds
\begin{equation}\label{eq3.24}
|E, \pm \rangle = \frac{ {\rm i}^{(1\mp 1)/2}\sqrt{m/ \hbar^2
    k}}{\big|1+T(k) + R(k)\big| }
\big(\hat\Omega_+ + \hat\Omega_-\big)\big( |k \rangle \pm
|-k\rangle\big)/\sqrt{2}~. 
\end{equation}
Using (\ref{eq3.22}) this can be
re-expressed by $T, R$, and $\hat\Omega_+$ as
\begin{equation}\label{eq3.25}
|E, \pm \rangle = {\rm i}^{(1 \mp 1)/2} \sqrt{\frac{m}{\hbar^2 k}} \frac{1 +
\bar{T} \pm \bar{R}}{|1+T \pm R|} \,\hat\Omega_+ \big(|k \rangle \pm
|-k \rangle \big)/\sqrt{2}~.
\end{equation}
Inserting this into (\ref{eq3.18}) and introducing $k$ as
integration variable, $E/\hbar = \hbar~k^2/2m$, 
one obtains an expression for $\hat{\Pi}_0$ with coherences between $|k
\rangle$ and $|- k\rangle$. For a free
particle in one dimension ($T=1, R=0,~\hat\Omega_+=\Eins$) one obtains 
\beq\label{eq3.25a}
\hat{\Pi}_0^{(V=0)} = \frac{1}{2\pi m}
\int_0^\infty\!\! dk\int_0^\infty\!\! dk^\prime
\sqrt{k k^\prime} \left\{|k\rangle \langle k^\prime |+
| -k \rangle \langle -k^\prime |\right\}
\eeq
so that in the
free case there are no coherences between positive and negative
momenta. \\[.3cm] 
\noindent
{\bf Example 2:} We consider the three-dimensional case of the
Hamiltonian
\begin{equation}\label{eq3.26}
\hat{H} = \frac{\hat{\bf P}^2}{2m} + V(\hat r)
\end{equation}
and assume a purely continuous spectrum. $\hat{H}$ is invariant under
rotations, reflection $\hat{\sigma}_1$ at the $x_2-x_3$ plane, i.e. $\phi
\to \pi - \phi$ in spherical coordinates, and under time reversal.  In
the following we assume 
only invariance of $\hat{T}$ and its probability distribution operator
under rotations and covariance under the combined action $\hat{\sigma}_1
\hat{\Theta}$; this means the invariance of $\hat{\Pi}_0$ under these
operations. Again this implies the existence of a unique time operator
and its probability distribution, as shown now. 

The eigenfunctions of $\hat{H}$ can be chosen in the form $f_{El}(r)
Y_{lm} (\vartheta, \phi)$, $f_{El}$ real. We denote these states by $|Elm
\rangle$, with the normalization
\begin{equation}\label{eq3.27}
\langle Elm|E^\prime l^\prime m^\prime \rangle = \delta (E -
E^\prime) \delta_{l l^\prime} \delta_{m m^\prime}~.
\end{equation}
The action of $\hat{\sigma}_1 \hat{\Theta}$ is given by
\begin{equation}\label{eq3.28}
\hat{\sigma}_1 \hat{\Theta} |E lm \rangle = (-1)^m |Elm \rangle~.
\end{equation}
Invariance under rotations and under $\hat{\sigma}_1 \hat\Theta$ implies that
$\hat{\Pi}_0$ can be written in the form 
\begin{equation}\label{eq3.29}
\hat{\Pi}_0 = \frac{1}{2\pi\hbar}\sum_i \int dE\,dE^\prime \sum_{lm} b_{iml}(E)\, |Elm \rangle
\langle E^\prime lm |\,b_{ilm}(E^\prime),
\end{equation}
with real functions $b_{ilm}$. In principle, the allowed set of $l$ and $m$ in
the sum might depend 
on $E,~ E^\prime$. However, if scattering theory can be applied this is
not so, as seen further below. Note the diagonality in $l,m$. The
condition for the functions $b_{ilm}$, arising from the normalization of the
probability, now reads
\begin{equation}\label{eq3.30}
\sum_i \left( b_{ilm} (E)\right)^2 = 1 \qquad \mbox{for each}~l,m~.
\end{equation}
Equations (\ref{3.12})-(\ref{eq3.15}) remain the same, only with $\alpha$
replaced by $lm$, and therefore minimality of $\Delta T$ implies
\begin{equation}\label{eq3.31}
\hat{\Pi}_0 = \frac{1}{2\pi\hbar}\sum_i \sum_{lm} \int dE ~dE^\prime |Elm \rangle \langle
E^\prime lm|,
\end{equation}
where $\langle {\bf x}|Elm\rangle = f_{El}(r)Y_{lm} (\vartheta, \phi)
$, with $f_{El}$ real. Thus also  in this example the time operator is
unique and there are no coherences between different $lm$ values. 

Using scattering theory, (\ref{eq3.31}) can again be
expressed by the M{\o}ller operators $\hat\Omega_{\pm}$ and  partial waves
shifts, $\delta_l$. We write $|Elm \rangle_0$ when $V\equiv 0$. Then
\begin{equation}\label{eq3.32}
  \langle {\bf x}|Elm\rangle_0 = \sqrt{\frac{2}{\pi}}
  \sqrt{\frac{m}{\hbar^2 k}}\, j_l (kr)\,Y_{lm} (\vartheta, \Phi)
\end{equation}
and, from rotation symmetry,
\begin{equation}\label{eq3.33}
\hat{S}\,|Elm \rangle_0 = e^{2{\rm i} \delta_l} |Elm\rangle_0~.
\end{equation}
$\hat{\Omega}_\pm$ commutes with rotations and $\hat{\sigma}$, and again
$\hat{\Theta}\, \hat{\Omega}_\pm \hat{\Theta} = \hat{\Omega}_\mp$~.
{}From $\hat{H}\, \hat{\Omega}_\pm = \hat{\Omega}_\pm \hat{H}_0$ it
follows that $(\hat{\Omega}_+ + 
\hat{\Omega}_-) |Elm \rangle_0$ is an eigenstate of $\hat{H},
\hat{L}^2, \hat{L}_3$, and also 
of $\hat{\sigma}_1 \hat{\Theta}$ with eigenvalue $(- 1)^m$. Therefore it
is a real multiple of $|Elm \rangle$. To calculate this multiple we use
(\ref{eq3.35}) and obtain
\begin{equation}\label{eq3.36}
(\hat{\Omega}_+ + \hat{\Omega}_-) |Elm \rangle_0 = \left( 1+ e^{\mp 2 {\rm i} \delta_l}
\right) \hat{\Omega}_\pm |Elm \rangle_0~.
\end{equation}
Taking scalar products one obtains, from the unitarity of
$\hat{\Omega}_\pm$,
\begin{equation}\label{eq3.37}
|Elm \rangle = \frac{1}{|1 + e^{2{\rm i} \delta_l}|}\, (\hat{\Omega}_+ + \hat{\Omega}_-)
|Elm \rangle_0
\end{equation}
and, from (\ref{eq3.36}),
\begin{equation}\label{eq3.38}
|Elm \rangle = \frac{1 + e^{\mp 2 {\rm i} \delta_l}}{|1 + e^{2{\rm i}
    \delta_l}|}\, \hat{\Omega}_\pm |Elm \rangle_0~.
\end{equation}
Inserting this into (\ref{eq3.31}) for $\hat{\Pi}_0$,   using $k$ as integration variable, where $E/\hbar = \hbar k^2/2m$,
and the states
\beq\label{eq3.39}
|klm\rangle_0 \equiv \hbar \sqrt{k/m}|Elm\rangle_0,
\eeq
which are normalized to $\delta (k - k^\prime) \delta_{ll^\prime}
\delta_{m m^\prime}$, then
\beqa 
\hat{\Pi}_0 &=& \hat{\Omega}_+ \frac{\hbar}{2\pi m} \sum_{lm} \int_0^\infty dk
\int_0^\infty dk^\prime
\nonumber\\
&& \sqrt{k k^\prime}\,
\frac{1 + e^{- 2{\rm i} \delta_l (k)}}{|1+ e^{2{\rm i} \delta_l (k)}|} \,\frac{1+
  e^{2{\rm i} \delta_l (k^\prime)}}{|1+ e^{2{\rm i} \delta_l (k^\prime)}|} |klm
\rangle_{0\,0} \langle k^\prime lm| \, \hat\Omega_+^\dagger.
\label{eq3.40}
\eeqa
The free particle case results if one puts $\Omega_+ = \Eins$
and $\delta_l = 0$.
\section{Application to arrival times}\label{arrival}
Kijowski \cite{Kij}  considered, in three dimensions, a free particle of mass $m$
coming in from $x_1 = - \infty$ and determined the distribution of times
of arrival at a plane perpendicular to the $x_1$ axis, e. g. at $x_1 = 0$,
by using physically motivated postulates. These were: covariance,
normalization, invariance under all {\em Galilei} transformations which
leave  the plane $x_1 = 0$ invariant, invariance under the combined
action of time reversal and 
space  reflection $\sigma$: ${\bf x} \to - {\bf x}$, and minimal variance.
His result, denoted here by $\hat{\Pi}_t^{Kij}$, for the probability
distribution of particles with only positive momenta in the $x_1$
direction and for arrivals at the plane $x_1 = 0$, 
can be written as
\beqa 
\hat{\Pi}_t^{Kij} &=& 
\frac{\hbar}{2\pi m} \int^\infty_{-\infty} dk_2 \,dk_3 \int_0^\infty dk_1 \int_0^\infty
dk_1^\prime
\nonumber\\
&&~~~~~~~~~~~~~~~~ e^{{\rm i}\hbar(k_1^2-k_1^{\prime
    ^2})t/2m}\, \sqrt{k_1 k^\prime_1}\, |k_1, k_2, k_3 \rangle \langle
k_1^\prime, k_2, k_3|~. 
\label{eq4.6}
\eeqa

In the following, the methods of the previous sections will be used to
generalize this result in two ways. First, invariance under Galilei
transformations can be replaced by  a weaker condition, and second,
the restriction to particles coming in from one side can be lifted. 
We assume covariance,  invariance under {\it   translations} which
leave the plane $x_1 = 0$ invariant, 
invariance under the space reflection $\sigma$: ${\bf x }\to -{\bf
  x}$, time reversal 
covariance, and minimal variance. These symmetry assumptions are taken
over from the classical case. Indeed, if one has  an
ensemble of free classical particles and reflects each
particle trajectory by  $\sigma$, then the resulting reflected
ensemble has  the
same distributions for arrivals at the plane $x_1 = 0$ as the original
ensemble. In addition, one can go over to the time reversed
trajectories, defined 
by $x_\theta(t) = x(-t)$.  Then, if a particle arrives at the plane at
time $t'$, the corresponding particle on the time reversed trajectory
will arrive there at time $-t'$. This implies that for the time
reversed ensemble the arrival time distribution is obtained from the
original one by replacing $t$ by $-t$. Note that the arrival
location, i.e. the plane $x_1 = 0$, is singled out here by the
required reflection invariance under $\sigma$.

Under the above assumptions it will now be shown that $\Pi^A_{f\,t}$, the operator
for the probability distribution  of times of arrivals at the plane $x_1$, is given by 
\beqa 
\hat{\Pi}_{f\,0}^A & =& \theta(\hat v_{1})\,|\hat v_{1}|^{1/2}\,
\delta(\hat x_1)\, |\hat v_{1}|^{1/2}\,\theta(\hat v_{1})
\nonumber\\
&~&~~+~\theta(- \hat v_{1})\,|\hat v_{1}|^{1/2}\,
\delta(\hat x_1)\, |\hat v_{1}|^{1/2}\,\theta(- \hat v_{1}),
\nonumber\\
\hat{\Pi}_{f\,t}^A & =& \exp
\{{\rm i}\hat H_0t/\hbar\}\,\hat{\Pi}_{f\,0}^A \exp\{-{\rm i}\hat H_0t/\hbar\},
\label{eq4.7aa}
\eeqa
where $\hat v_{1}$ and $\hat x_{1}$ are  components of the velocity
and position operator and  $\theta(x)$ is
the usual step function ($\theta(x)=0$ for $x<0$ and $\theta(x)=1$ for $x>0$).
Note that classically the first term on the right-hand side corresponds to $\theta(v_1)\,v_1\,
\delta(x_1)$ which classically gives the number of particles crossing the plane
$x_1$ per second from the left. Similarly for the second term, and
their sum corresponds to $|v_1|\, \delta(x_1)$. Thus in this case the arrival time
distribution operator  can be regarded as a particular   -- positive -- quantization of the classical current. The usual quantum
mechanical probability current is not necessarily positive, not even 
for states with only positive momenta in the $x_1$ direction \cite{backflow}.

Inserting in (\ref{eq4.7aa}) two complete sets of momentum
eigenstates and using
\beq  
\la{\bf k}|\delta(\hat x_1)|{\bf k}'\ra= \frac{1}{2\pi}
\delta(k_2 - k_2')\, \delta(k_3-k_3'),
\nonumber
\eeq
$\hat{\Pi}_{f\,t}^A$ can be written as 
\beqa 
&&\hat{\Pi}_{f\,t}^A = \frac{\hbar}{2\pi m}\int_{- \infty}^\infty dk_2 \, dk_3
 \int_0^\infty\! dk_1\int_0^\infty\! dk_1'e^{{\rm i}\hbar(k_1^2-k_1^{\prime
     2})t/2m}
\nonumber\\ 
&&\sqrt{k_1 k_1^\prime}
\big\{ |k_1,k_2,k_3\ra \langle k_1^\prime k_2k_3|
  + |-k_1,k_2,k_3 \rangle \langle -k_1^\prime, k_2, k_3 | \big\}. 
\label{eq4.7a}
\eeqa
For states with $k_1 > 0$ this reduces to (\ref{eq4.6}). 
Note the absence of coherences between positive and negative $k_1$
values in the above expression. For dimensions $d$ other than three an analogous result
holds. For $d=1$ it becomes, under a change $t \to -t$, identical
to the clock time expression in (\ref{eq3.25a}).

To show (\ref{eq4.7a}) we introduce the notation
\begin{eqnarray} 
|{\bf k},+ \rangle &=&\big(|k_1, k_2, k_3 \rangle + |-k_1, k_2, k_3
\rangle\big)/ \sqrt{2},
\nonumber\\
|{\bf k}, - \rangle  &=& {\rm i}\,\big(|k_1, k_2, k_3 \rangle - | - k_1,
k_2,k_3\ra\big) / \sqrt{2}, \label{eq4.8} 
\end{eqnarray}
which is a $\delta$-normalized basis and invariant under $\hat{\sigma} 
\hat{\Theta}$ since $\hat{\sigma} 
\hat{\Theta} |{\bf k} \rangle = |{\bf k} \rangle$. The latter 
also implies that $\langle {\bf k} |\hat{\Pi}^A_{f\,0} |{\bf k^\prime}
\rangle$ must be real.  One now has
\begin{eqnarray} \label{eq4.9}
\langle {\bf k}, + |\hat{\Pi}^A_{f\,0}| {\bf k}^\prime, + \rangle &=&
\langle {\bf k}, -|\hat{\Pi}^A_{f\,0}| {\bf k}^\prime,- \rangle = \langle {\bf
  k}|\hat{\Pi}^A_{f\,0}| {\bf 
  k}^\prime \rangle,
\nonumber\\ 
\langle {\bf k}, + |\hat{\Pi}^A_{f\,0}| {\bf k}^\prime, - \rangle &=& 0 ~.
\end{eqnarray}
Translation invariance in the $x_1 = 0$ plane gives the general form
\beq \label{eq4.10a}
\langle {\bf k}, + |\hat{\Pi}^A_{f\,0}| {\bf k^\prime}, + \rangle =
\delta (k_2 - 
k_2^\prime) \delta (k_3 - k^\prime_3) M (k_1, k_2, k_3, k_1^\prime),
\eeq
where, for fixed $k_2$ and $ k_3,~M(\cdot, k_2, k_3, \cdot)$ is a
positive definite 
kernel and one can write, with real functions $b_i (k_1; k_2 k_3)$,
\beq \label{eq4.10b}
M(k_1, k_2, k_3, k_1^\prime) = \frac{\hbar}{2\pi m}\sum_i b_i (k_1; k_2, k_3) b_i
(k_1^\prime; k_2, k_3)
\eeq
(cf.  \ref{Kij} for more mathematical details). From this and 
(\ref{eq4.9}) we  
obtain, with $|{\bf k^\prime}, \pm \rangle \equiv |k_1^\prime k_2, k_3, \pm
\rangle$,
\beqa 
\hat{\Pi}^A_{f\,0} &=&\frac{\hbar}{2\pi m} \sum_i \int_{- \infty}^\infty dk_2 \int_{-
  \infty}^\infty
dk_3 \int_0^\infty dk_1 \int_0^{\infty} dk^\prime_1
\nonumber\\
&&~~~~ b_i(k_1; k_2, k_3) \Big\{|{\bf k}, + \rangle \langle {\bf
  k}^\prime,+|
~+~
|{\bf k},- \rangle \langle {\bf k}^\prime,-|\Big\}\,b_i(k_1^\prime; k_2, k_3).
\label{eq4.11}
\eeqa
Note again the diagonality, this time in $k_2$ and $k_3$.
Since $\hat{\Pi}^A_{f\,0}$ commutes with translations and hence with the
momentum operators $\hat{P}_2$ and $\hat{P}_3$,
the normalization condition for the  probability distribution
\beq
\frac{1}{2\pi}\,\langle \psi|\, e^{{\rm i} \hat{H} t/\hbar}\,
\hat{\Pi}^A_{f\,0}\, e^{-{\rm i} \hat{H}t/\hbar}\,| \psi \rangle
\nonumber
\eeq
for states with $\langle {\bf k}|\psi\ra$ vanishing at $k_1 = 0$, gives
the condition
\beqa\label{eqcond1}
&\frac{\hbar}{ m}&\sum_i\int dk_1 dk_3 \int_0^\infty dk_1 dk_1^\prime
\,\delta \left(\frac{\hbar}{2m} (k_1^2 - k_1^{\prime 2})\right)
\nonumber \\ 
&\times& b_i(k_1; k_2, k_3)\,  b_i(k_1^\prime; k_2, k_3)
\left\{\overline{ \psi_+ (k_1 k_2 k_3)}\, \psi_+
(k_1^\prime k_2 k_3) + (+ \leftrightarrow -)  \right\}
\nonumber\\
& =& \sum_i \int dk_2~ dk_3 \int_0^\infty\!\! dk_1 \frac{1}{k_1} b_i(k_1;
k_2, k_3)^2 \left\{ |\psi_+|^2 + |\psi_-|^2 \right\}  
= 1, 
\eeqa
where $\psi_\pm ({\bf k}) \equiv  \langle {\bf k}, \pm |\psi \rangle $.
Since $|\psi \rangle$ can range through a dense set this yields
\beq \label{eq4.12}
\sum_i \left( \frac{b_i (k_1; k_2, k_3)}{\sqrt{k_1}}
\right)^2 = 1 ~, ~~~~k_1 >0~.
\eeq
For states $|\psi \rangle$ with $\psi_\pm({\bf k})$ vanishing at $k_1 =0$
one easily calculates that $\langle \psi |\hat{T}|\psi \rangle$ is
independent of $b_i$ and that for $\Delta T^2$ the only $b_i$
dependent term is of the form
\beq \label{eq4.13}
\int dk_2 \int dk_3 \int_0^\infty dk_1 \frac{1}{k_1} 
\sum_i\left(\partial_1 \frac{b_i}{\sqrt{k_1}}\right)^2 
\left\{ |\psi_+|^2 + |\psi_-|^2 \right\}
\eeq
where again the reality of $b_i$ is crucial. For $\Delta T^2$ to be
minimal this implies $\partial_1 (b_i/\sqrt{k_1}) = 0$ or
\beq \label{eq4.14}
\frac{1}{\sqrt{k_1}} b_i (k_1; k_2, k_3) = c_i (k_2, k_3)~.
\eeq
{}From (\ref{eq4.12}) it follows that  
\beq \label{eq4.15}
\sum_i (c_i (k_2, k_3))^2 = 1~.
\eeq
Inserting (\ref{eq4.14}) into (\ref{eq4.11}) and using
(\ref{eq4.15}) then gives
\beqa 
\hat{\Pi}^A_{f\,0} &=& \frac{\hbar}{2\pi m} \int_{- \infty}^\infty dk_2
\int_{- \infty}^\infty dk_3 
 \int_0^\infty dk_1 \int_0^\infty dk_1^\prime
\nonumber\\
&&\sqrt{k_1 k_1^\prime}
\left\{ |{\bf k},+ \rangle \langle k_1^\prime k_2k_3,+| + |{\bf k},-
  \rangle \langle k_1^\prime k_2 k_3, -| \right\}.\label{eq4.16}
\eeqa
Inserting for $|{\bf k},\pm \rangle$  from
(\ref{eq4.8}) one obtains the  expression in (\ref{eq4.7a}) for $t=0$.
The  operator $\hat{\Pi}^A_{f\,t}$ is then obtained from
(\ref{eq5.9a}). Since $\hat{\Pi}^A_{f\,0}$ commutes with $\hat
P_2$ and $\hat P_3$, only $\hat P_1^2$ remains and this yields
(\ref{eq4.7a}).

The absence of coherences  between
positive and negative $k_1$ components in (\ref{eq4.7a})  is due to the
assumed reflection invariance. Recall that invariance under 
Galilei transformations as in \cite{Kij} was not used here, only invariance under
translations of the plane $x_1 = 0$ is assumed.  

If instead of the plane 
$x_1 = 0$ one considers arrivals at the plane $x_1 = a$, the
corresponding probability distribution operator is obtained  by a spatial
translation, which leads to an additional  factor $\exp \left\{{\rm i}\,  (k_1 -
  k_1^\prime)a  \right\}$ in (\ref{eq4.7a}). 
\section{Interactions: conditional and operator-normalized arrival times }\label{cond}
In the derivation of the general form of covariant time operators the
normalization to 1 of the probability distribution  is an essential
condition. However, a free particle in two dimensions will arrive 
at  a finite interval with a probability less than 1, and similarly  in three
dimensions. For a particle in a potential, also the arrival
probability at a plane 
may be less than 1, due to scattering, and moreover, there may be
several arrivals. Experimentally, in such a case one may measure
the first time of arrival at a given position for a large number of replicas of
the system and determine their distribution with respect to the
total number of replicas. This distribution is, in general,  not 
normalized to 1 since there may be systems with no arrival
occurring. Dividing by the total probability   for a first arrival or,
equivalently,  considering only those systems for which an arrival has
actually been found, one obtains the
conditional first-arrival probability distribution. Although this is, by
construction, normalized to 1, it is not bilinear in the state vector and not  
the expectation of an operator. 
Alternatively, one may consider operator normalization as in
\cite{Fredenhagen,Seidel,HSMN} and then apply the techniques of
Section \ref{general}. 
The physical relevance and interpretation of operator-normalized distribution in terms of a modification of the initial state
has  been pointed out in \cite{HSMN}.

The operator for the (possibly non-normalized) distribution of first
arrivals is again denoted by   
$ \hat\Pi^{\,A}_{t}$ and the operator for the total probability for
arrival at a given location by $\hat N$. One has 
\beq \label{N1}
\hat N = \int_{-\infty}^{\infty}dt~ \hat\Pi^{\,A}_{t}\,.
\eeq
For a given state $|\psi\ra$ the conditional distribution, denoted by
$\Pi^{\,A\,\psi}_{\,c\,t}$ for the times of first arrivals is the given by 
\beq \label{c1}
\Pi^{\,A\,\psi}_{\,c\,t} = \frac{\la \psi| \hat\Pi^{\,A}_{t} |\psi \ra}{\la \psi| \hat{N}  |\psi \ra},
\eeq
which is evidently not bilinear in $|\psi\ra$.
Since  $\hat N$ is a positive operator,  $\hat N^{-1/2}$ exists on
states on which $\hat N$ does not vanish and which can be taken care
of by a projector.  If the external potential
remains finite or is not too singular one expects from tunneling that
for any state there is a finite arrival probability, i.e. that $\hat N
|\psi\ra$ will never vanish.
One can therefore define the operator $\hat\Pi^{\,A}_{N\,t}$ by
\beq \label{N2}
\hat\Pi^{\,A}_{N\,t} \equiv \hat N^{-1/2}\, \hat\Pi^{\,A}_{t} \hat
N^{-1/2}\,. 
\eeq
Its time integral is then the unit operator and therefore the distributions
determined by $\hat\Pi^{\,A}_{N\,t}$ are normalized to 1. If $\hat N$ and  $
\hat\Pi^{\,A}_{N\,t} $ are known one obtains  $ \hat\Pi^{\,A}_{t}$
from (\ref{N2}).

To the normalized operator distribution $ \hat\Pi^{\,A}_{N\,t} $ one can apply the
procedures of Section \ref{general} if additional symmetries hold. 
Physically motivated symmetries of  $ \hat\Pi^{\,A}_{t}$ 
carry over to  $\hat N$ and thus also over to
$\hat\Pi^{\,A}_{N\,t}$. More tricky is the requirement of minimal
variance under given symmetry conditions. Physically, one would
require minimal variance for the conditional distribution, i.e. that
$\hat\Pi^{\,A}_{t}$ is chosen in such a way that
$\Pi^{\,A\,\psi}_{\,c\,t}$ has minimal variance for each $|\psi\ra$ 
among all candidates satisfying the symmetry requirements. From
(\ref{c1}) one has
\beq \label{c2}
\Pi^{\,A\,\psi}_{\,c\,t} = \left\la \frac{\psi}{||\hat N^{1/2}|\psi\ra||}{\Big|}\,
\hat\Pi^{\,A}_{t}{\Big|} \frac{\psi}{|| \hat N^{1/2}|\psi\ra||} \right\ra .
\eeq
If one defines the normalized vector $|\psi_N\ra$ by
\beq \label{c4}
|\psi_N\ra \equiv \frac{\hat N^{1/2}|\psi\ra}{|| \hat N^{1/2}|\psi\ra||}
\eeq
one can write
\beq 
\Pi^{\,A\,\psi}_{\,c\,t} = \la \psi_N|\,\hat{N}^{-1/2}
\hat{\Pi}^{\,A}_{t}\,\hat N^{-1/2}|\psi_N \ra
=\la \psi_N|\, \hat{\Pi}^{\,A}_{N\,t}\,|\psi_N \ra .
\label{c5}
\eeq
For given $ \hat{N}$, the correspondence between $|\psi \ra $ and
$|\psi_N \ra $ is one to one and therefore, if  $ \hat\Pi^{\,A}_{t}$
leads to minimal variance the same is true for $\hat\Pi^{\,A}_{N\,t}$,
as long as $\hat N$ is kept fixed. Therefore, one may look for a
normalized operator distribution satisfying required symmetry
conditions and minimal variance, as in Section \ref{general}. If in
addition  the operator $\hat N$ for the total probability for arrival
at a given location is known, one then obtains the physically
interesting conditional probability distribution from Eqs. (\ref{N2})
and (\ref{c1}).  

To exemplify this, we consider one-dimensional motion in a bounded
potential. The latter  is assumed to be invariant under space  reflection about the
origin, and we consider an ensemble of particles which originally came in from $\pm
\infty$.  For an ensemble of classical particles there are two types
of individual trajectories. A particle coming in from -$\infty$
(+$\infty$) will either be reflected before it reaches the origin or,
if it reaches the origin, will then continue to +$\infty$ (-$\infty$),
by the symmetry of the potential. We  consider first arrivals at the
origin. The corresponding classical arrival time distribution has
certain symmetries. Indeed, from the reflection invariance of the potential,
in the ensemble obtained by space
reflecting each trajectory of the original ensemble has the same
distribution for arrivals at the origin as the original
ensemble. Going over to the time reversed trajectories  one concludes,
as in Section \ref{arrival} for the time
reversed ensemble, that the arrival time distribution is obtained from the
original one by replacing $t$ by $-t$. 

These symmetry properties are now carried over to the quantum case by
demanding corresponding symmetries for $\hat\Pi^{\,A}_{\,t}$,
i.e. invariance under space reflection and covariance under time
reversal. This implies the corresponding properties for the
operator-normalized $\hat\Pi^{\,A}_{N\,t}$ and in particular invariance of
$\hat\Pi^{\,A}_{N\,0}$   under space reflection and time
reversal. This then is the same situation as in Example 1 of Section
\ref{uniqueness} where a clock operator with the similar symmetry
properties was constructed. Therefore, for $\hat\Pi^{\,A}_{N\,0}$ on obtains
the same result as in (\ref{eq3.18}), where, as a reminder,
$|E,\pm \ra$  denote eigenstates of $\hat H$ such that $\la x|E,\pm
\ra$ are  real wavefunctions which are either symmetric or antisymmetric. 
>From this expression and  Eqs. (\ref{eq5.7a})~-~(\ref{eq5.8a})  one
then obtains $\hat{\Pi}^A_{N\,t}$. 

To determine the corresponding conditional probability distribution
$\hat N$ is needed, the operator for the total arrival
probability at the origin. An at least approximate expression for
$\hat N$  may  be obtained by modeling the
detection process. The simplest way to do this is by a complex
potential at the origin  or by a fluorescence model
(cf. \cite{toatqm2} for a review of these models). From  the $\hat N$ calculated
in this way and from the 
above result for $\hat{\Pi}^A_{N\,t}$ one then obtains an expression
for the conditional distribution of arrivals at the origin. The
complex potential and fluorescence models  also allows the
calculation of an approximate expression for the conditional arrival
distribution, and one may then compare the two expression, in particular
whether the  abstractly obtained operator-normalized
distribution  together with the 
approximate result for $\hat N$ leads to smaller variances than the
distributions derived from the models.

This example can be carried over to three dimensions and a Hamiltonian
of the form
\beq \label{eq5.1}
\hat{H}= \hat{\bf P}^2/2m + V(|\hat X_1|)~,
\eeq
using techniques similar to those described in Appendix \ref{Kij}.

\section{Discussion}
We have derived the most general form of covariant  time operators
with a normalized probability distribution. The result has been
applied to clock time operators and to time-of-arrival operators. To
arrive at a unique operator, the multitude of possible operators has
been restricted by physically motivated symmetry requirements,
in particular time reversal invariance and,  to obtain an optimally
sharp observable, by the requirement of minimal variance.

Our general result has allowed us not only to weaken the assumptions
of Kijowski \cite{Kij} but at the same time generalize his
arrival-time distribution for free particles arriving at a 
plane. It had been required in  \cite{Kij} that the distribution
should be invariant  under Galilei transformations which transform the
plane into itself. As shown here,  Galilei transformations are not
needed and  translation invariance is sufficient. Moreover, 
the restriction that the particles should come in from one side of the
plane only has been lifted.

A free particle in three dimensions will arrive 
at a location of finite extent only with a probability less than 1,
and also for a particle in a potential the arrival probability at a plane
may be less than 1, due to scattering, and moreover, there may be
several arrivals. In an actual and repeated repeated measurement one can
consider only those outcomes for which at least one arrival has
actually been found at some time and then in addition consider only the 
distribution of the times of the first arrival. This amounts to a
division by the total arrival probability and determines the
(normalized) conditional first-arrival probability distribution. 
However, the latter is not bilinear in the state vector, due to the division. 
As an alternative, we have applied operator normalization to the original
distribution, and then one can apply the general results. The distribution
thus obtained leads to the conditional distribution if the total
arrival probability is known. 


It should be pointed out that the probability  distribution associated
with an observable is
in general of more direct experimental relevance than the 
operator of the observable itself because in most circumstances first
the distribution of 
individual measurement results for a given state is obtained and then
the expectation value is derived from that. For a particle in a
potential, the total arrival probability at a plane may be less than
one. In such a case there may still be a non-normalized temporal probability
distribution operator but not necessarily an associated time
operator. As has been pointed out above, the conditional probability
distribution is in general not bilinear in the state vector and so cannot be used
directly to construct a time operator.

In order to be meaningful, any experimental procedure purported to be the
realization of a measurement of arrival times or of a quantum clock
has to satisfy certain requirements, and the so constructed ideal
observables provide guide-lines for these requirements. In this context, an
analysis of this kind for the
ideal arrival time-of-arrival distribution  
of \cite{Kij} has been carried out in terms of an operational 
quantum-optical realization with cold atoms (cf.  \cite{toatqm2} for a review).

\section*{Acknowledgments}
We thank Michael J. W. Hall for discussions. 
We acknowledge the kind hospitality of the Max Planck 
Institute for Complex Systems in Dresden, as well as  
funding by the Basque Government (Grant IT472-10), and the  
Ministerio de Ciencia e Innovaci\'on (FIS2009-12773-C02-01).   

\appendix
\section{Mathematical details}   
\subsection{Nondegenerate energy spectrum}\label{nondegenerate}
In this appendix, mathematical details are given to justify some more
formal arguments in the main text. To illustrate the crucial points we
first treat the case of a Hamiltonian $\hat{H}$ with a simple
absolutely continuous spectrum from $[E_0, \infty)$ or $(- \infty,
\infty)$, with no degeneracy. The generalized eigenvectors can then be
denoted by $|E \rangle$, with normalization $\langle E|E^\prime
\rangle = \delta (E- E^\prime)$. In the Hilbert space, $\cal{H}$, of
the Hamiltonian we denote by ${\cal S}(\hat {H})$ and ${\cal{D}}(\hat{H})$
the subspaces of vectors $|\psi\rangle \in {\cal{H}}$ such that $\langle E|\psi
\rangle \equiv \psi(E)$ is in the Schwartz space ${\cal S}(\mathbb{R})$
or $\cal{D}(\mathbb{R})$, respectively. Then $\hat{\Pi}_0 = -{\rm i}
[\hat{H}, \hat{F}_0]/\hbar$ from (\ref{eq1.7bc}) is a positive
semi-definite continuous sesquilinear form on $ {\cal S}(\hat{H})$ or
${\cal{D}} (\hat{H})$ and, by continuation, on  $ {\cal
  S}(\mathbb{R})$ or $ {\cal D}(\mathbb{R})$.
Hence, by the kernel theorem, it defines a
distribution on ${\cal S}(\mathbb{R}^2)$ or ${\cal{D}}(\mathbb{R}^2)$,
respectively, formally denoted by
\beq\label{eqA.1}
\Pi_0(E, E^\prime) \equiv \langle E|\hat{\Pi}_0| E^\prime \rangle~.
\eeq
Let $\left\{ |g_i\rangle \in {\cal{D}} (\hat H)~,~ i = 1,\cdots \right\}$ be
maximal set (finite or infinite) such that
\beq \label{eqA.2}
\langle g_i |\hat{\Pi}_0| g_j \rangle = \delta_{ij}~.
\eeq
We put $g_i(E)\equiv \la E|g_i\ra$ and define $b_i \in {\cal
  S}'(\mathbb{R})$ by 
\beq \label{eqA.3}
b_i(E) =\sqrt{2\pi\hbar} \int dE^\prime\, \Pi_0(E, E^\prime)\, g_i(E^\prime)~.
\eeq
Then
\beq \label{eqA.4}
\langle g_\alpha | b_i \rangle \equiv \int dE \,\overline{g_\alpha (E)}
 \, b_i(E) = \sqrt{2\pi\hbar}~\delta_{\alpha i}
\eeq
and, therefore,
\beq \label{eqA.5}
\sum_i \langle g_\alpha|b_i \rangle \langle b_i |g_\beta \rangle = 2\pi\hbar~
\delta_{\alpha \beta}~.
\eeq
With ${\cal{L}} \{g_i\}$ the linear span of the
$g_i(E)\,$'s one therefore has, on ${\cal{L}} \{g_i\}
\times {\cal{L}} \{g_i\}$,  
\beq \label{eqA.6}
\langle E  | \hat{\Pi}_0 | E^\prime \rangle  =\frac{1}{2\pi\hbar}
\sum_i b_i(E) \overline{ b_i(E')}~. 
\eeq
If ${\cal{N}} \subset {\cal{S}} (\mathbb{R})$ or $\subset{\cal{D}} (\mathbb{R})$
defines the null space of (the distribution defined by) $\hat{\Pi}_0$,
then ${\cal{L}} \{g_i\} \cup {\cal{N}}$ span ${\cal S}(\mathbb{R})$ or
${\cal{D}} ({\mathbb{R}})$, respectively, and therefore, by continuity,
(\ref{eqA.6}) holds also on ${\cal S}(\hat H) \times {\cal S}(\hat
H)$ and ${\cal{D}}(\hat H) \times {\cal{D}}(\hat H)$, respectively.\\

\noindent
{\bf Exploiting normalization}. From Eqs. (\ref{eq1.7b}) and
(\ref{eqA.6}) one obtains, for a normalized $|\psi \rangle \in
 {\cal S}(\hat H)$, 
\begin{eqnarray} 
1 &=& \int \frac{dt}{2 \pi\hbar} \langle\psi | e^{-{\rm i} \hat{H}t/\hbar} \hat{\Pi}_0\,e^{{\rm i} \hat{H}t/\hbar}|\psi 
\rangle \nonumber \\ 
&=& \sum_i \int d\tau \left|
\frac{1}{\sqrt{2 \pi}} \int dE \,e^{{\rm i} E\tau} \,\overline{b_i (E)}\, \psi (E)
\right|^2.
\label{eqA.7}
\end{eqnarray}
This implies that, for each $i$, the inner integral, considered
as a function of $\tau$, is square integrable, and so is its Fourier
transform. The latter, as a distribution, is given by $\overline{b_i
  (E)}\, \psi(E)$ and hence is equivalent to a square integrable
(i.e. $L^2$) function. Hence $b_i(E)$ is not only a distribution but even
locally an $L^2$-function. From 
(\ref{eqA.7}) one then has by Parseval's theorem
\beq \label{eqA.8}
\sum_i \int dE\, \left| b_i(E)\, \psi (E) \right|^2 = 1~.
\eeq
Since this is true for any normalized $|\psi\rangle \in {\cal S}({\hat H})$,
this implies 
\beq \label{eqA.9}
\sum_i \overline{b_i(E)}\, b_i(E) = 1 ~~\mbox{for almost all}~E.
\eeq
Conversely, if a sequence of locally $L^2$-functions satisfies
(\ref{eqA.9}), then $\hat{\Pi}_0$ defined by (\ref{eqA.6}) gives
rise to a covariant time operator through Eqs. (\ref{eq1.7b}) and
(\ref{eq1.2}).\\

\noindent
{\bf Exploiting the second moment.} We assume that for a dense set of
$|\psi\rangle$ in ${\cal S}({\hat H})$ the second moment  exists, i.e.
\beq \label{eqA.10}
\int \frac{dt}{2 \pi\hbar}\, t^2\, \langle \psi| e^{-{\rm i} \hat{H}t/\hbar} \hat{\Pi}_0\, e^{ {\rm i}
  \hat{H}t/\hbar}|\psi \rangle \,< \, \infty ~.
\eeq
With (\ref{eqA.6}) this can be written as
\beq \label{eqA.11}
\hbar^2\sum_i\int d\tau\, \tau^2\, \left| \frac{1}{\sqrt{2 \pi}}   \int dE\, e^{{\rm i}
E \tau } \,\overline{b_i(E)}\, \psi (E) \right|^2~.
\eeq
This implies that, for each $i$, the inner integral considered as a
function of $\tau$ is not only square integrable but is also in the domain
of the operator ``multiplication by $\tau$''. Therefore, by standard
results, the Fourier transform of this function, given by
$\overline{b_i (E)} \psi (E)$, is not only in $L^2$ but  is also
absolutely continuous and has a derivative which is also in $L^2$. Again by
Parseval's theorem one has
\beq \label{eqA.12}
\mbox {second moment} =\hbar^2 \sum_i \int dE \left|
\partial_E \left( \overline{b_i(E)}\, \psi (E)  \right)\right|^2 < \infty~.
\eeq
Since $|\psi \rangle$ is from the dense set  ${\cal S}({\hat H})$, $b_i$
must also be 
absolutely continuous and its derivative exists almost everywhere and
is locally $L^2$. Using (\ref{eqA.9}) one obtains
\beqa 
&&\mbox {second moment}/\hbar^2
=
\\ 
&&\int\! dE
\left| \partial_E \psi
\right|^2 + \int \!dE \sum_i |b_i^\prime \, \psi|^2
 + 2\,{\rm Re}\, \int\! dE \sum_i
\overline{b_i^\prime}\, b_i\, \bar{\psi^\prime}\, \psi
\label{eqA.13}
\eeqa
In a similar way one finds
\beq \label{eqA.14}
\langle \psi |\hat{T}| \psi \rangle = \int dE\, \bar{\psi}\, i\hbar \,\partial_E
\psi +  \int dE\, |\psi|^2 \sum b_i\,{\rm i}\hbar\, \partial_E\overline{b_i}~.
\eeq
We note that if the $b_i\,$'s are real then (\ref{eqA.9}) implies
$\sum  b_i(E)\, b_i^\prime (E) = 0$ for almost all $E$ and thus for real
$b_i\,$'s the last term in (\ref{eqA.13}) as well as in
(\ref{eqA.14}) vanishes.
\subsection{Extension to degenerate energy spectrum.} \label{degenerate}
In case of a discrete degeneracy label we use the normalization 
as in (\ref{eq2.7})
where for simplicity we assume the same degeneracy for each $E$. (The
general case can be treated by direct integrals of Hilbert
spaces). Then $\Pi(E,E')$, $g_i (E)$ and $ b_i(E)$ from the
nondegenerate case are  replaced by $\Pi(E,\alpha,E',\alpha ')$, 
$g_i (E, \alpha)$ and $ b_i(E,\alpha)$, respectively, and
(\ref{eqA.6}) can be written as 
\beq \label{eqA.15}
\langle E ,\alpha | \hat{\Pi}_0 | E^\prime, \alpha^\prime \rangle =
\frac{1}{2 \pi\hbar}\sum_i 
  b_i(E, \alpha)\, \overline{b_i (E^\prime, \alpha^\prime)}~. 
\eeq
By the same technique as above the normalization condition now implies
\beq \label{eqA.16}
\sum_i b_i (E, \alpha)\, \overline{b_i (E, \alpha^\prime)} = \delta_{\alpha
  \alpha^\prime}~.
\eeq
Finiteness of the second moment implies that $b_i (E, \alpha)$ is
absolutely continuous in $E$ for each $\alpha$ and its derivative is
locally $L^2$. 

In case of a continuous degeneracy label we use the normalization
\beq \label{eqA.17}
\langle E, \lambda |E^\prime \lambda^\prime \rangle = \delta
(E-E^\prime)\, \delta (\lambda - \lambda^\prime)
\eeq
This can be reduced to the discrete case by choosing a complete
orthonormal set $\left\{\Phi_\alpha (\lambda),~ \alpha = 1,2, \cdots
\right\}$ and putting
\beq \label{eqA.18}
| E, \alpha \rangle \equiv \int d \lambda\, \Phi_\alpha (\lambda)\,| E,
\lambda \rangle ~,~ \alpha = 1,2, \cdots
\eeq
It can also be treated directly by replacing the functions $g_i(E)$ by
$g_i (E, \lambda)$ and $b_i(E)$ by $b_i(E, \lambda)$. Then
(\ref{eqA.16}) is replaced by the equation
\beq \label{eqA.19}
\sum_i b_i(E, \lambda)\, \overline{b_i (E, \lambda^\prime)} = \delta
(\lambda - \lambda^\prime)
\eeq
valid for almost all $E$.

\subsection{The case of partial translation invariance}\label{Kij}

To illustrate some mathematical details for Section
\ref{arrival}  we consider for notational simplicity the
two-dimensional case.   We search for a covariant 
probability distribution operator, 
 denoted by $\hat \Pi_t^A$, which is invariant under
 reflection $\sigma: {\bf x} \to -{\bf x}$, under translations which
 leave the line $x_1 =0$ invariant, as well as  under time reversal
$\hat{\Theta}$, and which has minimal
variance. Let $\hat H_1$ be the one dimensional Hamiltonian
\beq \label{eqA.19b}
\hat{H}_1 = \hat P^2_1/2m 
\eeq
in the Hilbert space ${\cal{H}}_1 \equiv L^2 ({\mathbb{R}})$ and
denote by $|E_1, \pm \rangle$ the real symmetric resp. antisymmetric
eigenstates of 
$\hat{H}_1$, similarly as in Example 1 of Section \ref{uniqueness}. The
eigenstates of $\hat{H}$ are then
\beq  \label{eqA.19c}
|E_1, \pm \rangle \,| k_2 \rangle~.
\eeq
The assumed invariances imply the corresponding invariances of
$\hat{\Pi}^A_0$ and of $\hat{F}^A_0$. The expectation value of the
latter gives the 
probability of finding a measured time value in $(- \infty, 0)$, and
so $0 \leq \hat{F}_0^A  \leq 1$. If $|\psi \rangle \in {\cal H}$, the
formal expression
\beq \label{eqA.19d}
| \psi (k_2)\ra \equiv \langle k_2 | \psi \rangle
\eeq
can be regarded, for each $k_2$, as a vector in ${\cal H}_1$, i.e. as a
vector function of $k_2$ with values in ${\cal{H}}$, and for $|\psi_1
\rangle,\, |\psi_2 \rangle \in {\cal{H}}$ one has
\beq \label{eqA.19e}
\int dk_2 \langle { \psi}_1 (k_2)| { \psi}_2 (k_2) \rangle_{{\cal H}_1 } =
\langle \psi_1 |\psi_2 \rangle
\eeq
where the index indicates the scalar product in ${{\cal H}_1 }$.
Each bounded operator $\hat{A}$ which commutes with $\hat{P}_2$
satisfies
\beq \label{eqA.19f}
\langle k_2 |\hat{A} |k'_2 \rangle = \delta (k_2 - k_2^\prime)\,
\hat{A}(k_2)~, 
\eeq
where $\hat{A}(k_2)$ acts in ${\cal H}_1$, and thus
\beq \label{eqA.19g}
\langle k_2| \hat{A}| \psi \rangle = \hat{A} (k_2) |\psi (k_2)\ra~.
\eeq
In a rigorous way this can be expressed in terms of direct integrals of
Hilbert spaces,
\beq \label{eqA.19h}
{\cal{H}} = \int_\oplus d k_2\, {\cal{H}}( k_2) ~,\qquad \mbox{where}~~
{\cal{H}}(k_2) \equiv {\cal{H}}_1~,
\eeq
with $| \psi \rangle$ corresponding to a vector-valued function 
$|\psi(\cdot)\ra$, 
with $|\psi(k_2)\ra \in {\cal{H}} (k_2)$, and $\hat{A}$
corresponding to an operator-valued function $\hat{A}(\cdot)$
satisfying (\ref{eqA.19g}).
If $\hat{A}$ is positive and bounded by 1 then so is $\hat{A}(k_2)$,
for almost all $k_2$.

We now take $\hat{F}_0^A$ for $\hat{A}$. Since
$\hat{H}_1$ acts in each ${\cal{H}} (k_2)$, the action of
$\hat{\Pi}_0^A$ is given by
\beq \label{eqA.19j}
\hat{\Pi}_0^A(k_2) \equiv \frac{- {\rm i}}{\hbar} \left[ H_1,~\hat F_0^A (k_2)  \right]
\eeq
considered as a bilinear form in each ${\cal{H}} (k_2)$. For almost all $k_2$,
$\hat{\Pi}_0^A(k_2)$ is a 
positive bilinear form satisfying invariance under $\hat\sigma$ and
$\hat\Theta$. Hence each $\hat{\Pi}_0^A(k_2)$ is given by (\ref{eq3.18}),
(with $E$ replaced by $E_1$). This finally implies
\beq \label{eqA.19k}
\hat{\Pi}_0^A = \frac{1}{2 \pi \hbar}\int d E_1 \,dE_1^\prime \,\Big\{
|E_1, + \rangle \langle E_1, +|\,+\,  |E_1,- \rangle \langle E_1^\prime, - |
\Big\}
\eeq
or, equivalently,
\beqa 
\langle \psi |\hat{\Pi}_0^A| \psi \rangle &=& \frac{1}{2 \pi
\hbar}\int dk_2\,dE_1\,dE_1^\prime
\nonumber\\ 
&&\Big\{ \langle \psi |E_1 + \rangle |k_2 \rangle \langle k_2 |\langle
  E_1^\prime, + |\psi \rangle ~+~ \langle + \leftrightarrow - \rangle
\Big\}.
\label{eqA.19l}
\eeqa
In three dimensions, the procedure is analogous. Now (\ref{eq4.16}) results by a change of variable. 
\section*{References}

\end{document}